\documentclass{book}    % fundamental class file is 'book'.
\usepackage{piers}  % use the file 'piers.sty'.
\begin{document}

\title{ Fast Computation of Electromagnetic Wave Propagation and Scattering
for Quasi-cylindrical Geometry   } \maketitle

\begin{authors}

{\bf Shaolin Liao}   \\
%\medskip
Electrical and Computer Engineering, 1415 Engineering Drive, Univ.
of Wisconsin, Madison, U.S.A., 53706
\end{authors}
%--------------------------
%---Content of Paper Abstract-----------------------
\begin{paper}

\begin{piersabstract}
The cylindrical Taylor Interpolation through FFT (TI-FFT) algorithm
for computation of the near-field and far-field in  the
quasi-cylindrical geometry has been introduced. The modal expansion
coefficient  of the vector potentials ${\bf F}$  and ${\bf A}$
within the context of the cylindrical harmonics (TE and TM modes)
can be expressed in the closed-form expression through the
cylindrical addition theorem. For the quasi-cylindrical geometry,
the modal expansion coefficient can be evaluated through FFT with
the help of the Taylor Interpolation (TI) technique. The near-field
on any arbitrary cylindrical surface can be obtained
  through the Inverse Fourier Transform (IFT).  The far-field   can  be obtained
  through the Near-Field Far-Field (NF-FF)
transform.  The cylindrical  TI-FFT algorithm has the advantages of
$ \mathcal{O} \left( \hbox{N} \log_2 \hbox{N} \right) $
computational complexity for $\hbox{N} = \hbox{N}_\phi \times
\hbox{N}_z$ computational grid, small sampling rate (large sampling
spacing) and no singularity problem.
\end{piersabstract}

  \begin{center}
  {\bf \small I. INTRODUCTION}
\end{center}
The planar Taylor Interpolation  through FFT (TI-FFT) algorithm introduced before \cite{Shaolin} has been shown to be efficient in
the computation of  narrow-band beam propagation and scattering for the quasi-planar geometry \cite{liao_image_2006, liao_near-field_2006, shaolin_liao_new_2005, liao_fast_2006,  liao_beam-shaping_2007, liao_fast_2007, liao_validity_2007, liao_high-efficiency_2008, liao_four-frequency_2009, vernon_high-power_2015, liao_multi-frequency_2008, liao_fast_2007-1, liao_sub-thz_2007, liao_miter_2009, liao_fast_2009, liao_efficient_2011, liao_spectral-domain_2019}.  However, cylinder-like geometry is not uncommon in the electromagnetic engineering, e.g.,  the  input mirror system design \cite{liao_sub-thz_2007} for the high-power  gyrotron application. In such case, the planar TI-FFT algorithm is not efficient and we have developed the cylindrical TI-FFT to solve the problem. 

 For the cylindrical geometry, the computation is efficient because  the electromagnetic field that is
expressed in the cylindrical harmonics can be numerically
 implemented through the  FFT. For the quasi-cylindrical geometry,
the FFT can still be used, with the help of  the Taylor
Interpolation (TI) technique. Fig. \ref{scheme} shows the scheme
used to illustrate the cylindrical TI-FFT algorithm and the time
dependence $e^{i \omega t}$  ($i \equiv \sqrt{-1}$)is used in this
article.

\begin{center}
  {\bf \small II. THE NEAR-FIELD AND THE FAR-FIELD }
\end{center}

In this section, the near-field and the far-field for surface
currents (${\bf M}_s$, ${\bf J}_s$) are presented within the context
of  the cylindrical harmonics.

\begin{center}
  {\bf \small 1.  The Near-field  }
\end{center}

It can be shown \cite{liao_validity_2007} that the vector potential (${\bf F}$, ${\bf A}$)  due to  surface currents (${\bf M}_s$, ${\bf J}_s$) for the scattering phenomenon in the region $\rho > \rho'$ can be expressed as 
\begin{eqnarray}\label{FA}
 \begin{array}{cc} {\bf F} ({\bf r}) \\ {\bf A} ({\bf r}) \end{array}    =
    \hbox{IFT} \left\{ \  \begin{array}{cc}  {\bf f}_m^h  \\  {\bf g}_m^h  \end{array}
      H_m^{(2)} ( \Lambda \rho ) \
 \right\},
\end{eqnarray} 

  \begin{eqnarray}\label{fg}
  \begin{array}{cc}  {\bf f}_m^h  \\  {\bf g}_m^h  \end{array}
   =  \frac{ 1}{i4}
     \int \! \int_{S} dS' \   \begin{array}{cc} \epsilon  {\bf M}_s {\bf (r')}  \\  \mu  {\bf J}_s {\bf (r')}  \end{array}
     H_m^{(1)} ( \Lambda \rho )  e^{i m  \phi'}   e^{ i h z'},
\end{eqnarray}
where  $H_m^{(1)} ( \ {\cdot} \ )$ and $H_m^{(2)} (
\ {\cdot} \ )$ are Hankel functions of the first kind and the second
kind of integer order $m$ respectively.  The Inverse Fourier
Transform (IFT) has been defined as,
\begin{equation}\label{IFT}
\hbox{IFT} \left\{ \hspace{-0.14in} \begin{array}{cccc}  \\
\end{array}  \   {\bf \cdot} \  \right\}  =   \frac{1}{2 \pi}
\sum_{m=-\infty}^\infty \int_{-\infty}^{\infty} d
 h       \left\{  \hspace{-0.14in} \begin{array}{cccc}  \\
\end{array} \   {\bf \cdot}  \  \right\}  e^{ - i m  \phi}      e^{-i h
 z}.
\end{equation}

The electromagnetic field (${\bf E}$, ${\bf H}$) is given as 
\begin{equation}\label{E}
 {\bf E}( {\bf r}) =   -\frac{1}{\epsilon} \nabla \times {\bf F} ({\bf r}) - i
 \omega {\bf A} ( {\bf r}) + \frac{1}{i \omega \epsilon \mu} \nabla'
 \left[ \hspace{-0.14in} \begin{array}{cccc}  \\ \\
\end{array} \nabla' \cdot   {\bf A}  ( {\bf r}) \right].
\end{equation} 

\begin{equation}\label{H}
 {\bf H}( {\bf r}) =   \frac{1}{\mu} \nabla \times {\bf A} ({\bf r}) - i
 \omega {\bf F} ( {\bf r}) + \frac{1}{i \omega \epsilon \mu} \nabla'
 \left[ \nabla' \cdot   {\bf F}  ( {\bf r}) \right].
\end{equation} 

\begin{figure}[t]\centering
\includegraphics[width=3.4 in, height= 2.6 in]{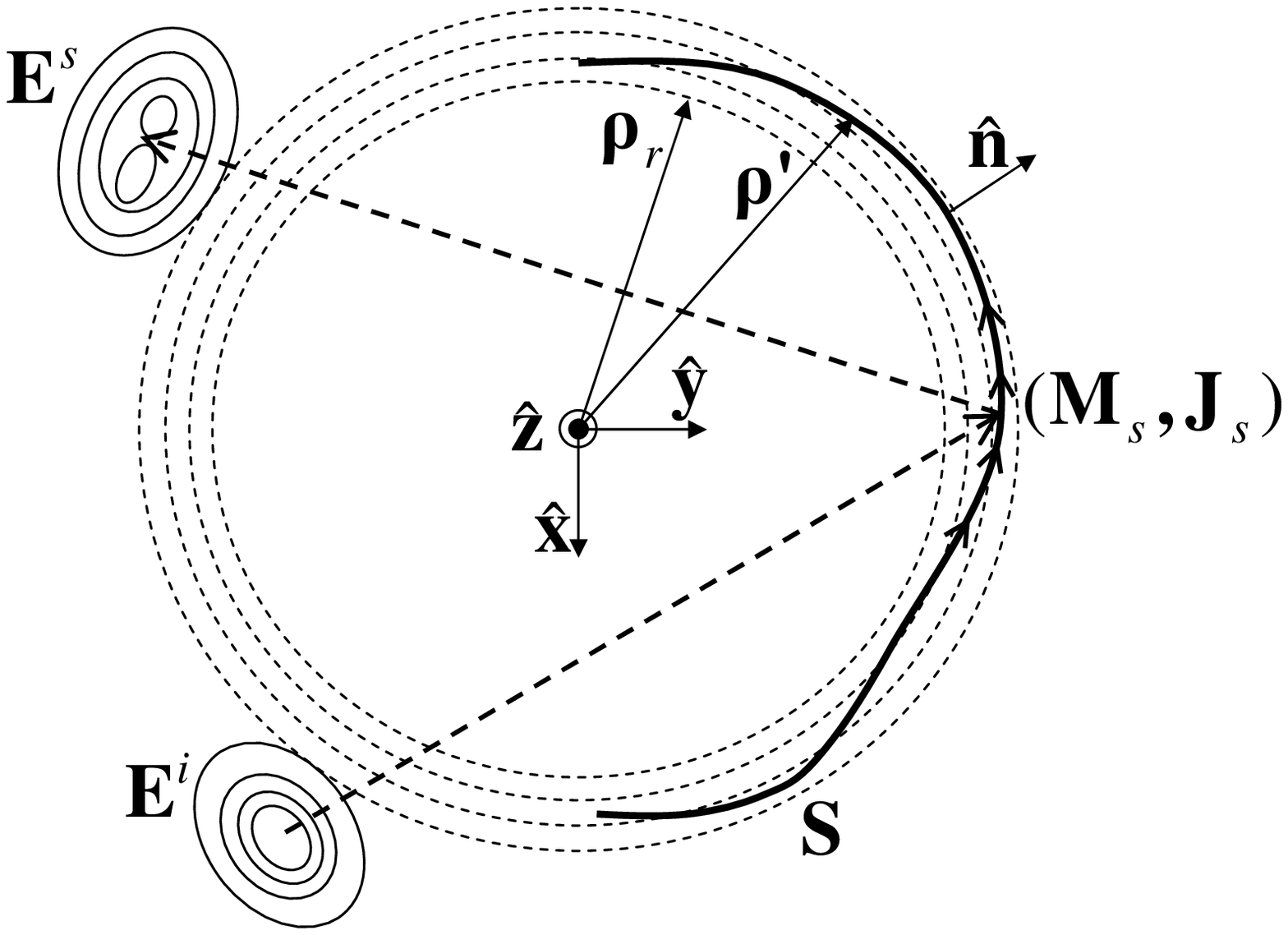}
\caption{The scattering of the narrow-band beam: the incident field
${\bf E}^i$ propagates onto PEC surface $S$ and is back-scattered to
${\bf E}^s$. The induced surface currents (${\bf M}_s, {\bf J}_s$)
can be obtained through the Method of Moment (MoM) or Physical
Optics (PO) approximation if PEC surface $S$ is smooth enough.
$\rho'$ is the source coordinate and $\rho_r$ is the radius of the
reference cylindrical surface. $\hat{\bf n}$ is the surface normal
to $S$.
 }
 \label{scheme}
\end{figure}
 
 \begin{center}
  {\bf \small 2.  The Cylindrical Harmonics  }
\end{center}
 
The cylindrical TE  and TM modes are obtained when the magnetic
(electric) surface current has only $\hat{\bf z}$-component, i.e.,
${\bf M}_s = \hat{\bf z} \hbox{ M}_{s,z}$ (${\bf J}_s = \hat{\bf z}
\hbox{ J}_{s,z}$). From (\ref{FA})-(\ref{H}),

\begin{eqnarray}\label{M}
   {\bf M}^{ h }_{m } ( {\bf r})   = \left[  \hspace{-0.07in}
\begin{array}{ccc}  \\  \\  \end{array}
   \hat{\boldsymbol{\rho}} \frac{m}{i\rho} H_m^{(2)}( \Lambda \rho)
   -   \hat{\boldsymbol{\phi}} \Lambda  \frac{\partial H_m^{(2)}( \Lambda \rho)}{\partial (\Lambda
   \rho)}  \hspace{-0.07in}
\begin{array}{ccc}  \\  \\  \end{array} \right] e^{-i m\phi} e^{- i h z}.
\end{eqnarray}
\begin{eqnarray}\label{N}
   {\bf N}^{ h }_{m } ( {\bf r})   =
 \left[  \hspace{-0.07in}
\begin{array}{ccc}  \\  \\  \end{array}  \hat{\boldsymbol{\rho}} \frac{h \Lambda}{ik} \frac{\partial H_m^{(2)}( \Lambda \rho)}{\partial (\Lambda \rho)}
   -   \hat{\boldsymbol{\phi}} \frac{mh}{k\rho} H_m^{(2)}( \Lambda
   \rho) + \hat{\bf z} \frac{\Lambda^2}{k} H_m^{(2)}( \Lambda \rho) \hspace{-0.07in}
\begin{array}{ccc}  \\  \\  \end{array} \right] e^{-i m\phi} e^{- i
h z}.
\end{eqnarray}

The electromagnetic field (${\bf E}$, ${\bf H}$) can be expressed as
the combination of the TE and TM modes,

%\vspace*{-0.1in}

\begin{eqnarray}\label{ETETM}
 { \bf E}  ( \rho )      =      \sum_{m}
 \left\{
\int_{-\infty}^{\infty}    \left[  \hspace{-0.07in}
\begin{array}{ccc}  \\  \\  \end{array}  a^{h}_{m} \ {\bf M}^{
h }_{m} (\rho )   +  b^{ h}_{m}  \ {\bf N}^{h }_{m}
 (\rho ) \hspace{-0.07in} \begin{array}{ccc}  \\  \\  \end{array} \right] dh
  \hspace{-0.07in} \begin{array}{ccc}  \\  \\ \\  \end{array}
  \right\},
  \end{eqnarray}

\begin{eqnarray}\label{HTETM}
 { \bf H}  ( \rho )      =
\frac{i}{\eta}  \sum_{m}
 \left\{
\int_{-\infty}^{\infty}    \left[  \hspace{-0.07in}
\begin{array}{ccc}  \\  \\  \end{array}  a^{h}_{m} \ {\bf N}^{
h }_{m} (\rho )   +  b^{ h}_{m}  \ {\bf M}^{h }_{m}
 (\rho ) \hspace{-0.07in} \begin{array}{ccc}  \\  \\  \end{array} \right] dh
  \hspace{-0.07in} \begin{array}{ccc}  \\  \\ \\  \end{array}
  \right\},
  \end{eqnarray}

\begin{eqnarray}\label{ab}
 a^{h}_{m}  =  - \frac{1}{ 2 \pi \epsilon}     \hbox{   f}_{m,z}^h, \ \ \ \      b^{ h}_{m}
 =  - \frac{i v}{ 2 \pi}   \hbox{   g}_{m,z}^h,
  \end{eqnarray}

\hspace{-0.2in}where $\eta = \sqrt{\frac{\mu}{\epsilon}}$ and $v =
\frac{1}{\sqrt{\mu \epsilon}}$ is the electromagnetic wave velocity
in the homogeneous medium. 

\newpage

 \begin{center}
  {\bf \small 3.  The Far-field }
\end{center}

The far-field can be obtained through the Near-Field Far-Field
(NF-FF) transform \cite{Leach},

\begin{eqnarray}\label{Efar}
 { \bf E}  ( {\bf R} )       =     - \frac{2 k \sin \theta e^{-i k R}}{R} \sum_{m}
    i^m e^{-im \phi}  \left[  \hspace{-0.07in}
\begin{array}{ccc}  \\  \\  \end{array} \hat{\boldsymbol{\phi}}
a^{h}_{m}
  +  \hat{\boldsymbol{\theta}} i   b^{ h}_{m}   \hspace{-0.07in} \begin{array}{ccc}  \\  \\  \end{array}
 \right],
  \end{eqnarray}

\begin{eqnarray}\label{Hfar}
 { \bf H}  ( {\bf R} )       =      - \frac{2 k \sin \theta e^{-i k R}}{ \eta R} \sum_{m}
    i^m e^{-im \phi}  \left[  \hspace{-0.07in}
\begin{array}{ccc}  \\  \\  \end{array} \hat{\boldsymbol{\phi}}  i   b^{ h}_{m}
  -  \hat{\boldsymbol{\theta}} a^{h}_{m}   \hspace{-0.07in} \begin{array}{ccc}  \\  \\  \end{array}
 \right],
  \end{eqnarray}
where ${\bf R}$ is the coordinate in the far-field
and $R = \left| {\bf R} \right|$.

\begin{center}
  {\bf \small III. THE CYLINDRICAL TI-FFT ALGORITHM}
\end{center}

For the narrow-band beam and the quasi-cylindrical surface, both the
electromagnetic field in (\ref{ETETM})-(\ref{HTETM}) and the modal
expansion coefficient in (\ref{fg}) can be expressed in the Taylor
series, which facilitates the use of FFT.   Due to the similarity,
only TE mode will be considered in this article.

\begin{center}
  {\bf \small 1. The Electromagnetic Field}
\end{center}

Generally, the near-field  ${\bf E}$ can be expressed in the Taylor
series,

\vspace{-0.1in}

\begin{eqnarray}\label{ETI}
 \hbox{\bf E}(\rho_r+ \delta \rho) = \left.\hbox{\bf E}(\rho_r) +
\sum_{n=1}^\infty \frac{1}{n!} \frac{\partial^{(n)} {\bf E}}{
\partial \rho^{(n)} }\right|_{  \rho_r}   (\delta \rho)^n,
\end{eqnarray}
where, $\rho_r$ is the reference cylindrical surface
and the Taylor coefficient  $\left.\frac{\partial^{(n)} {\bf
E}}{\partial \rho^{(n)} }\right|_{  \rho_r}$ can be expressed in the
form of IFT. Take TE mode ($ {\bf M}^{ h }_{m }$) as an example, for
$\hat{\boldsymbol{\phi}}$-component $\hbox{E}_{\phi}$,  from
(\ref{M})  and  (\ref{ETETM}), 

\begin{eqnarray}\label{Ephi}
 \hbox{E}_{\phi}({\bf \rho})  = \hbox{IFT}  \left\{ \ \frac{\Lambda}{\epsilon}
 \frac{\partial H_m^{(2)}(\Lambda \rho)}{\partial (\Lambda
 \rho)}   \hbox{   f}_{m,z}^h \  \right \},
\end{eqnarray}

Now, the Taylor coefficient  for $\hbox{E}_{\phi}({\bf \rho})$ are
given as 

\begin{eqnarray}\label{CoefEphi}
\left. \frac{\partial^{(n)} {\bf E}}{\partial \rho^{(n)} }\right|_{
  \rho_r} = \hbox{IFT}  \left\{ \ \frac{\Lambda^{n+1}}{\epsilon}
 \left. \frac{\partial^{(n+1)} H_m^{(2)}(\Lambda \rho )}{\partial (\Lambda
 \rho )^{(n+1)}} \right|_{\rho_r}   \hbox{   f}_{m,z}^h \  \right\}.
\end{eqnarray}

Similar argument  holds for other electromagnetic field components
and TM mode.

\begin{center}
  {\bf \small 2. The Modal Expansion Coefficient}
\end{center}

 Similarly, $H_m^{(1)}(\Lambda \rho')$ in the  the modal expansion coefficients (${\bf f}_m^h, {\bf g}_m^h$)
 in (\ref{fg}) can be expanded  into the Taylor series, 

\begin{equation}\label{Hm1TI}
 H_m^{(1)}\left(\Lambda\left[ \hspace{-0.1in} \begin{array}{ccc}  \\  \\    \end{array} \rho_r
 + \delta \rho' \hspace{-0.1in} \begin{array}{ccc}  \\   \end{array} \right]
 \right)   =  H_m^{(1)}(\Lambda \rho_r ) +
\sum_{n=1}^\infty \frac{\Lambda^n}{n!}  \left. \frac{\partial^{(n)}
H_m^{(1)} (\Lambda \rho) }{
\partial (\Lambda
\rho)^{(n)} }\right|_{ \rho_r}  (\delta \rho')^n
\end{equation}
where $\delta \rho' = \rho' - \rho_r$. Now, the modal
expansion coefficient in (\ref{fg}) is given as,

\begin{equation}
  \begin{array}{cc}  {\bf f}_m^h  \\  {\bf g}_m^h  \end{array} =     \sum_{\triangle S}
    \sum_{n=0}^\infty      \hbox{FT}  \left\{ \ \left.\gamma_m(\Lambda
 \rho_r)  \begin{array}{cc}  \epsilon  \tilde{\bf M}_{s} {\bf
(r')}  \\  \mu  \tilde{\bf J}_{s} {\bf (r')}  \end{array} (\delta
\rho')^n \ \right\}\right|_{\triangle S}
\end{equation}
where the Fourier Transform $\hbox{FT}$ is defined
similarly as $\hbox{IFT}$  in (\ref{IFT}) and $\triangle S$ is the
small surface patch between two adjacent reference cylindrical
surfaces; what's more, the following quantities have been defined,

\begin{eqnarray} \label{MJapp}
 \gamma_m(\Lambda \rho_r) =
 \frac{  \pi   }{i 2}    \left.\frac{\Lambda^n \rho'}{n!} \frac{\partial^{(n)}   H_m^{(1)} (\Lambda
\rho) }{\partial (\Lambda \rho)^{(n)} }\right|_{  \rho_r}, \ \ \ \
  \begin{array}{cc}  \tilde{\bf M}_{s} {\bf (r')}  \\  \tilde{\bf J}_{s} {\bf (r')}  \end{array}
    = \frac{1}{ \hat{\bf
n}  \cdot \hat{\boldsymbol{\rho}'}} \begin{array}{cc}   {\bf M}_{s}
{\bf (r')}
\\   {\bf J}_{s} {\bf (r')}   \end{array}.
\end{eqnarray}
$\hat{\bf n}$ is the surface normal to $S$.

 \begin{figure}
\begin{minipage}[t]{.5\textwidth}
 \includegraphics[width= 3 in, height= 2.3
in]{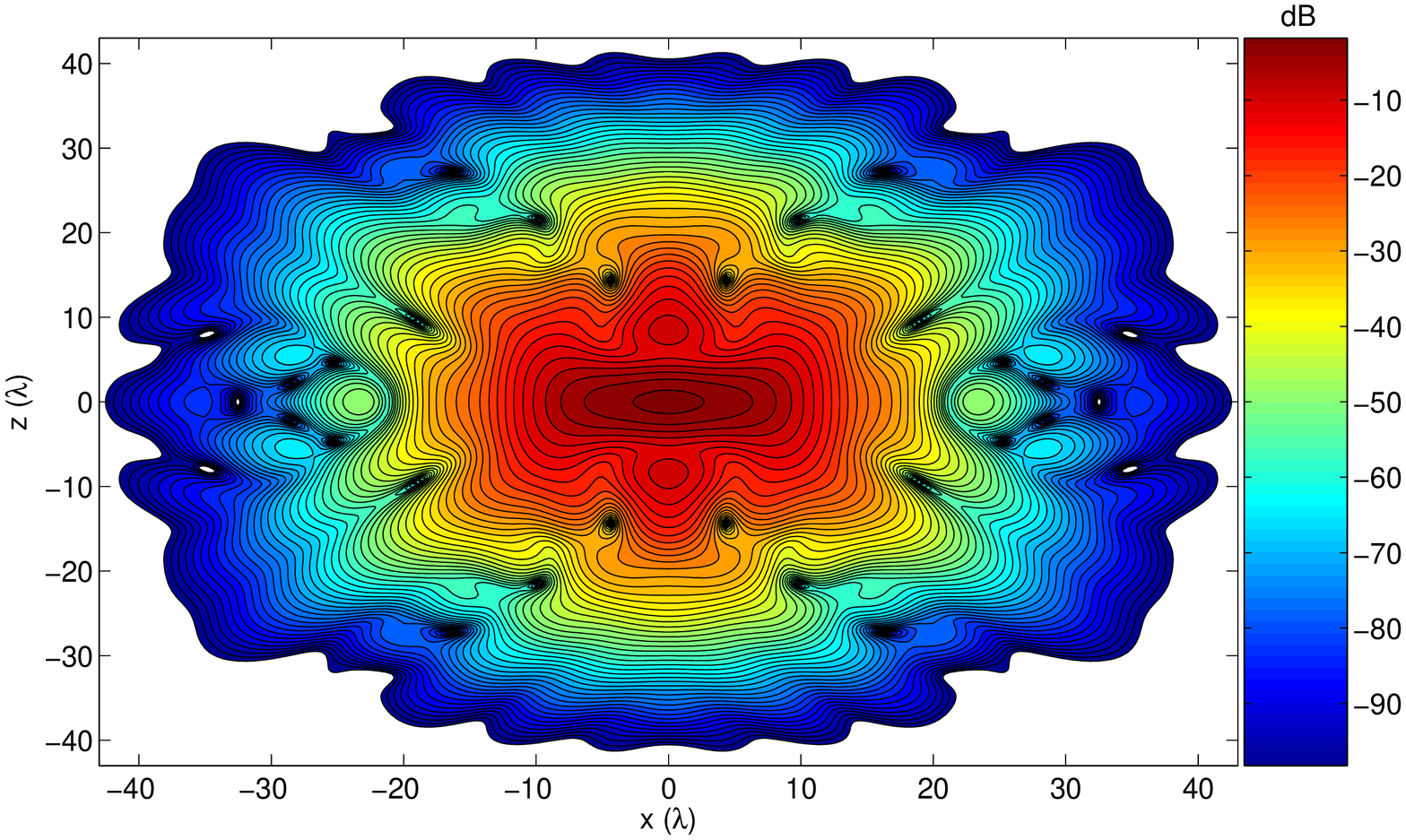}  \caption{ $20 \log_{10} |\hbox{E}_x^s|$.}
\label{fig:Ex}
\end{minipage}%
\begin{minipage}[t]{.5\textwidth}
 \includegraphics[width= 3 in, height= 2.3
in]{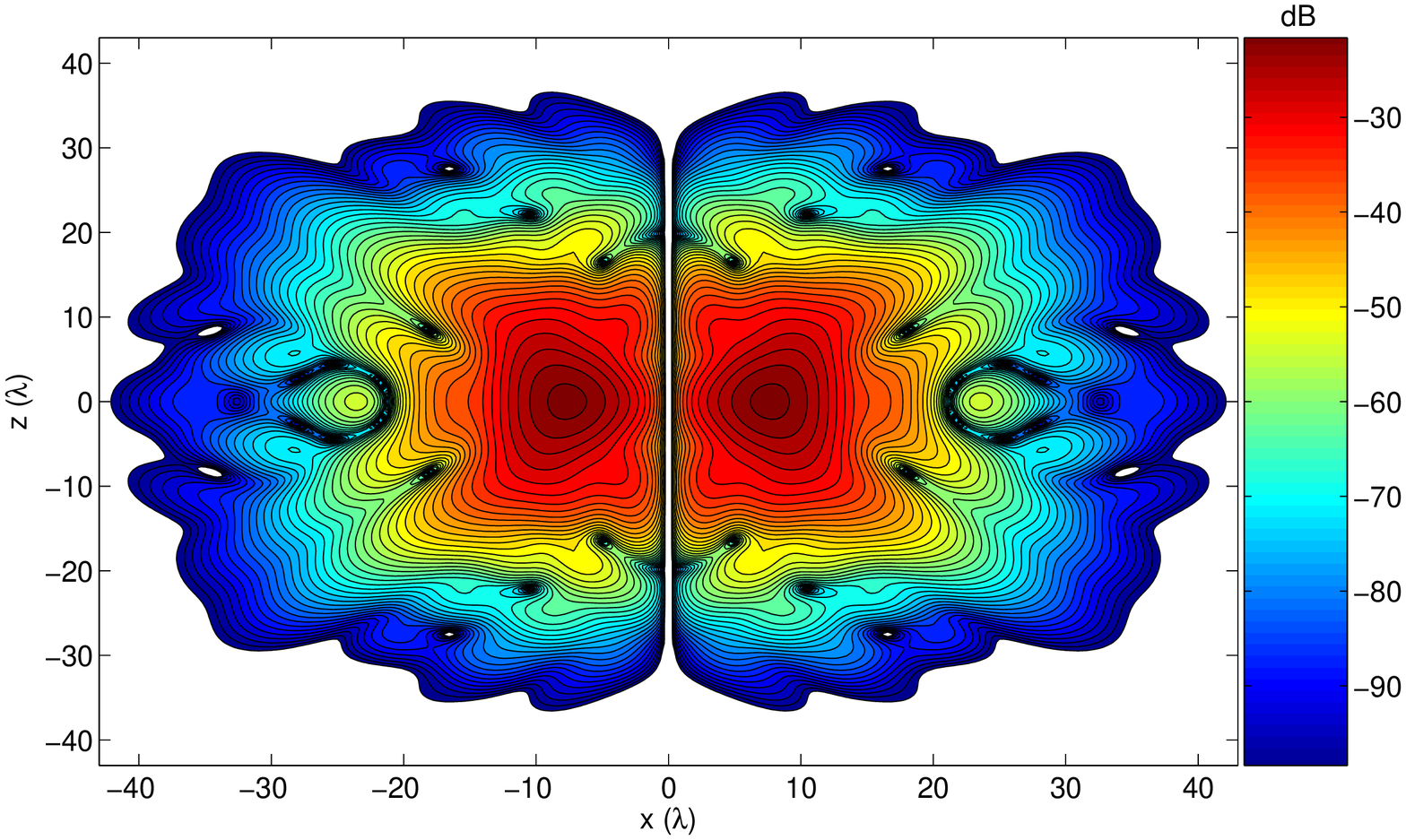}  \caption{ $20 \log_{10} |\hbox{E}_z^s|$.}
\label{fig:Ez}
\end{minipage}%
\end{figure}

 \begin{figure}[b]\centering
\includegraphics[width= 5.  in, height= 3.5  in]{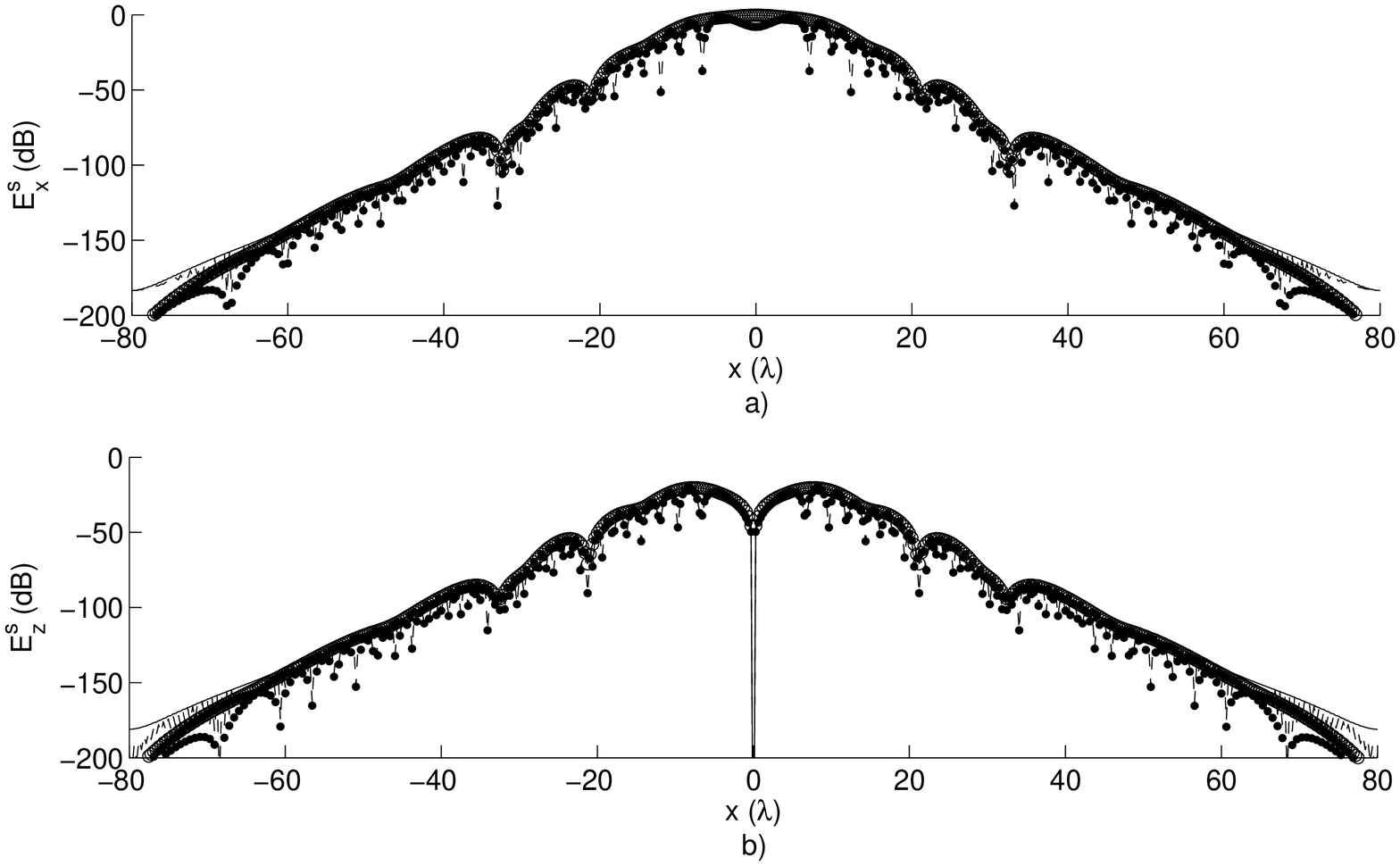}
\caption{Comparison of the cylindrical TI-FFT algorithm  with the
direct integration method: a) $\hbox{E}_x$ (dB); and b) $\hbox{E}_z$
(dB). Plots are shown in $\hat{\bf x}$ direction, across the maximum
value point of $|\hbox{E}_x|$. Solid and dashed lines denote the
magnitude and real part obtained from the direct integration method
respectively; circles and dots denote the magnitude and real part
obtained from the cylindrical TI-FFT algorithm respectively.
$\hbox{E}_y$ is small and not shown.} \label{fig:Eplots}
\end{figure}

  \begin{center}
  {\bf \small IV. NUMERICAL RESULT }
\end{center}

To show the efficiency of the cylindrical TI-FFT algorithm, the
direct integration  method \cite{Shaolin} has been used to make
comparison with the cylindrical TI-FFT algorithm. The numerical
example used for such purpose is a 110 GHz ($\lambda \sim$ 2.7 mm)
Fundamental Gaussian Beam (FGB) scattered by a PEC quasi-cylindrical
surface with a cosine wave perturbation. The incident  FGB is
$\hat{\bf x}$-polarized and  propagates at $\hat{\bf z}$ direction,
with symmetrical beam waist radii  $ w_x = w_y = 8  \lambda$. The
quasi-cylindrical PEC surface  is given as

\begin{eqnarray}
 y(x,z)    =   \sqrt{\left( 80 \lambda \right)^2 -
x^2}   +   0.1 \lambda \cos\left(2 \pi \frac{x}{ 20 \lambda}\right)
\cos\left(2 \pi \frac{z}{20 \lambda}\right)
\end{eqnarray} 

\begin{eqnarray}
\rho (x,z) = x \cos \phi + y \sin \phi, \ \ \ \ \phi = \arctan\left[
\frac{y}{x} \right] \nonumber
\end{eqnarray}

The scattered  field ${\bf E}^s$ is evaluated  on plane $y=0$ (where
the incident  FGB starts to propagate). Fig. \ref{fig:Ex} and Fig.
\ref{fig:Ez} show the magnitude patterns of the x-component
 $\hbox{E}_x^s$  and the z-component  $\hbox{E}_z^s$  of the scattered
output field ${\bf E}^s$ (y-component $\hbox{E}_y^s$ is small and
not shown). The comparison of  result obtained from the cylindrical
TI-FFT algorithm and that from the direct integration method is
given in Fig. \ref{fig:Eplots}, for both the magnitude and the real
part.

 The CPU time for the cylindrical TI-FFT algorithm t$_{\hbox{\tiny TI}}$ and the CPU time for
the direct integration method t$_{\hbox{\tiny DI}}$ are shown in
Fig. \ref{fig:cpu}.  The ratio t$_{\hbox{\tiny DI}}/$t$_{\hbox{\tiny
TI}}$ is shown in Fig. \ref{fig:cpucmp}, for different size of the
computational grid ($\hbox{N} = \hbox{N}_\phi \times \hbox{N}_z$).
All work was done in Matlab 7.0.1, on a 1.66 GHz PC, with Intel Core
Duo and 512 MB RAM.

 \begin{figure}
\begin{minipage}[t]{.5\textwidth}
\includegraphics[width= 3.  in, height= 2.1in]{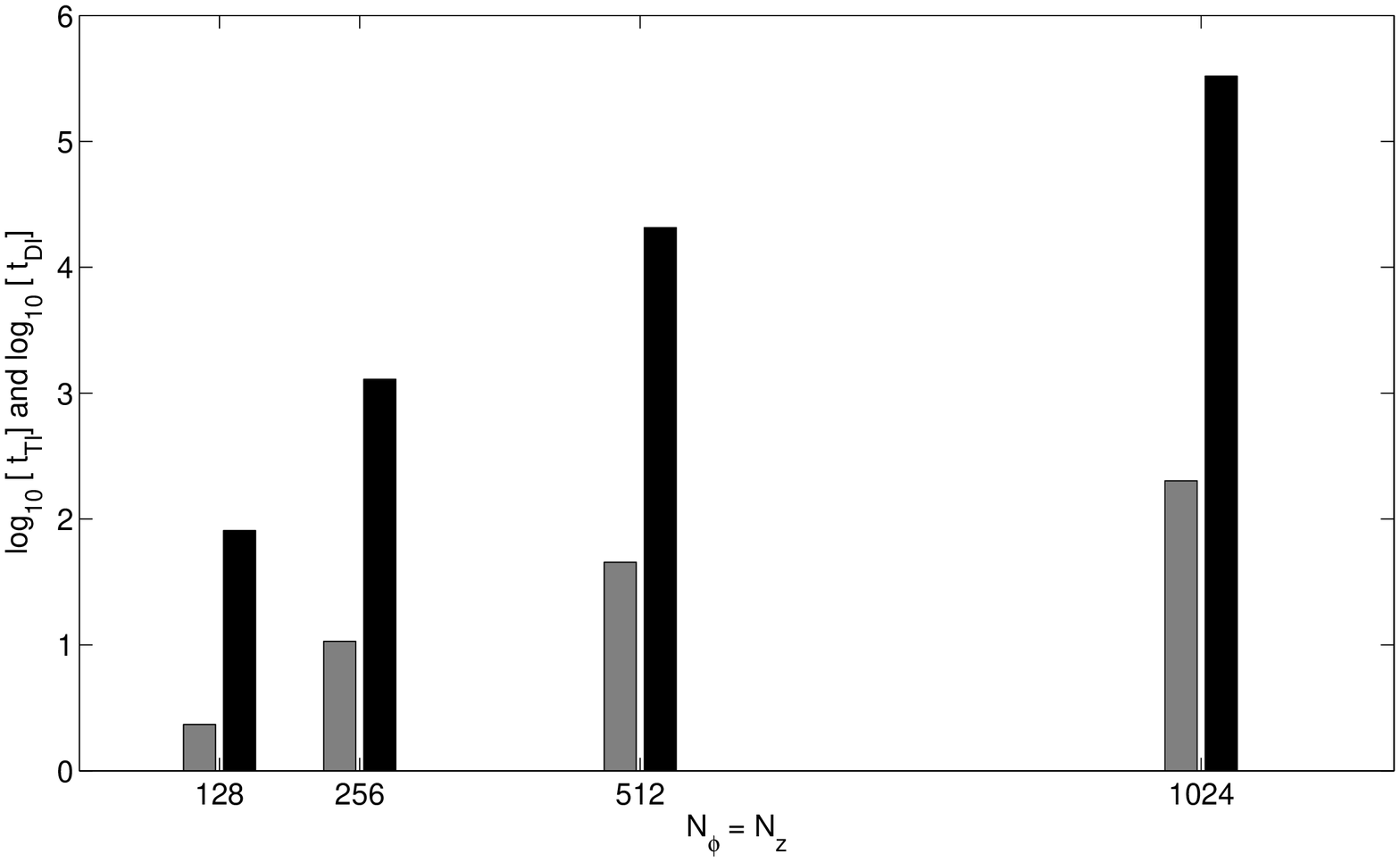}
\caption{ The logarithmic  CPU time   t$_{ \hbox{\tiny  TI}}$ and
t$_{ \hbox{\tiny DI}}$.}
 \label{fig:cpu}
\end{minipage}%
\begin{minipage}[t]{.5\textwidth}
\includegraphics[width= 3. in, height= 2.1 in]{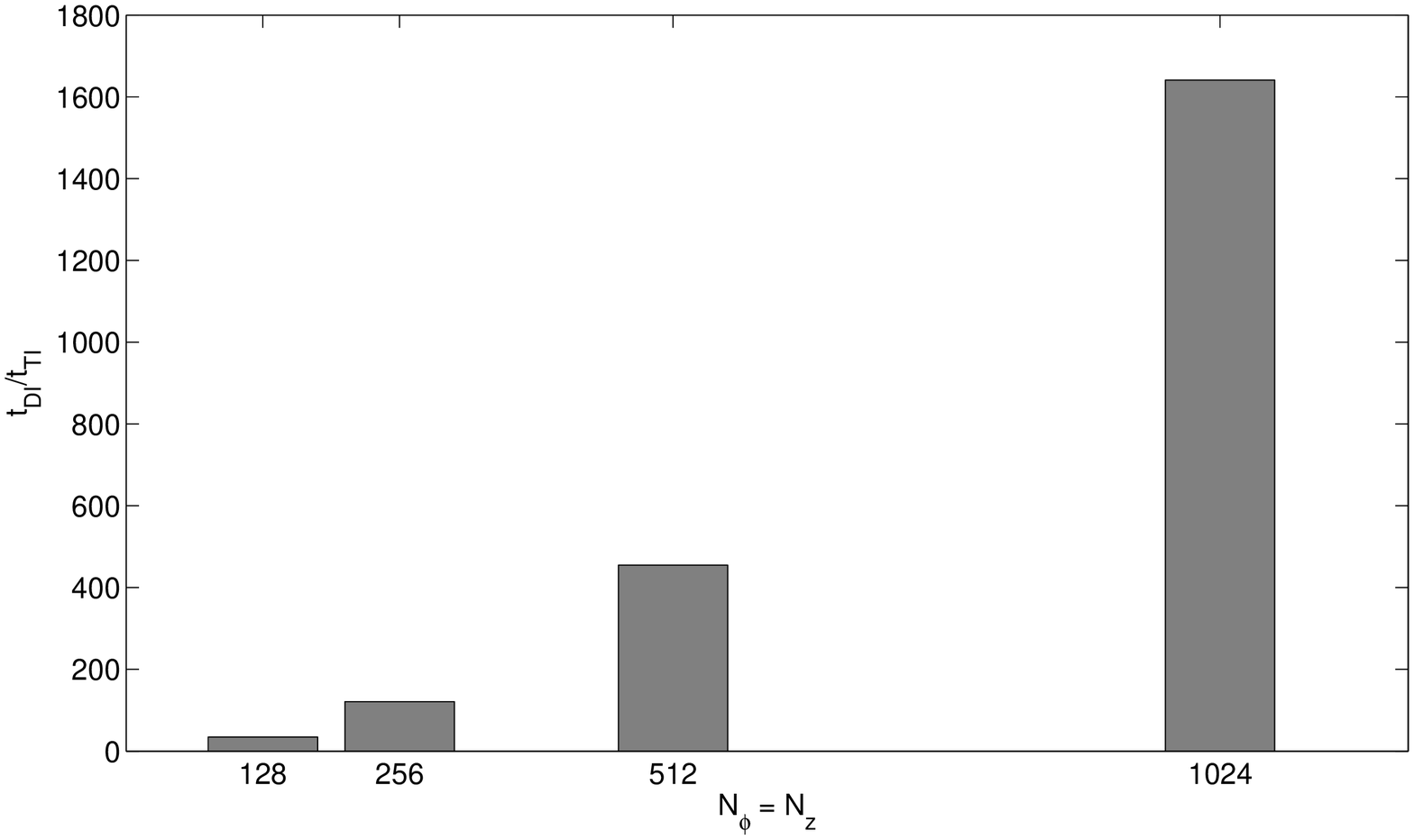}
\caption{ The CPU time ratio of t$_{ \hbox{\tiny DI}}/$t$_{
\hbox{\tiny TI}}$. } \label{fig:cpucmp}
\end{minipage}%
\end{figure}

  \begin{center}
  {\bf \small IV. CONCLUSION }
\end{center}

The cylindrical TI-FFT algorithm for the computation of the
electromagnetic wave propagation and scattering  has been introduced
for the narrow-band beam and the quasi-geometry geometry. The
cylindrical TI-FFT algorithm has the complexity of  $\mathcal{O}
\left( \hbox{N} \log_2 \hbox{N} \right)$ for $ \hbox{N} =
\hbox{N}_\phi \times \hbox{N}_z$ computational grid. The algorithm
allows for a low sampling rate (limited the Nyquist sampling rate)
and doesn't have the problem of singularity.

  \begin{center}
  {\bf Acknowledgements }
\end{center}

This work was supported by the U.S. Dept. of Energy under the
contract DE-FG02-85ER52122.

\end{paper}

\end{document}